\documentclass[english]{revtex4}
\usepackage[T1]{fontenc}
\usepackage[latin9]{inputenc}
\usepackage{amstext}
\usepackage{esint}

\usepackage{graphicx}

\makeatletter
\@ifundefined{textcolor}{}
{%
 \definecolor{BLACK}{gray}{0}
 \definecolor{WHITE}{gray}{1}
 \definecolor{RED}{rgb}{1,0,0}
 \definecolor{GREEN}{rgb}{0,1,0}
 \definecolor{BLUE}{rgb}{0,0,1}
 \definecolor{CYAN}{cmyk}{1,0,0,0}
 \definecolor{MAGENTA}{cmyk}{0,1,0,0}
 \definecolor{YELLOW}{cmyk}{0,0,1,0}
 }

\makeatother

\usepackage{babel}

\begin{document}

\title{A PNJL Model for Adjoint Fermions with Periodic Boundary Conditions}

\author{Hiromichi Nishimura}
\email{hnishimura@physics.wustl.edu}
\author{Michael C. Ogilvie}
\email{mco@wuphys.wustl.edu}
\affiliation{%
Department of Physics, Washington University, St.\ Louis, MO 63130, USA
}%

\date{\today}

\begin{abstract}
Recent work on QCD-like theories has shown that the addition of adjoint
fermions obeying periodic boundary conditions to gauge theories on
$R^{3}\times S^{1}$ can lead to a restoration of center symmetry
and confinement for sufficiently small circumference $L$ of $S^{1}$.
At small $L$, perturbation theory may be used reliably to compute
the effective potential for the Polyakov loop $P$ in the compact direction.
Periodic adjoint fermions act in opposition to the gauge fields, which
by themselves would lead to a deconfined phase at small $L$. In order
for the fermionic effects to dominate gauge field effects in the effective
potential, the fermion mass must be sufficiently small.
This indicates
that chiral symmetry breaking effects are potentially important. We
develop a Polyakov-Nambu-Jona Lasinio (PNJL) model which combines
the known perturbative behavior of adjoint QCD models at small $L$
with chiral symmetry breaking effects to produce an effective potential
for the Polyakov loop $P$ and the chiral order parameter $\bar{\psi}\psi$.
A rich phase structure emerges from the effective potential. Our results
are consistent with the recent lattice simulations of Cossu and D'Elia,
which found no evidence for a direct connection between 
the small-$L$ and large-$L$ confining regions.
Nevertheless, the
two confined regions are connected indirectly if an extended field
theory model with an irrelevant four-fermion interaction is considered. Thus
the small-$L$ and large-$L$ regions are part of a single confined phase.
\end{abstract}


\maketitle

\section{Introduction}

Recent progress in the study of QCD-like gauge theories has revealed
that a confined phase can exist under certain conditions
when one or more spatial directions
are compactified and small \cite{Unsal:2007vu,Myers:2007vc}.
This is surprising, because a small compact direction in Euclidean
time gives rise to a deconfined phase for $SU(N)$ gauge theories.
It is also intriguing, because one or more small compact directions give
rise to a small effective coupling constant if
the theory is asymptotically free. 
Thus we now have four-dimensional field theories in which confinement
holds, and holds under circumstances where semiclassical methods may
be reliably applied. In addition, the existence of confinement with
one or more compact directions suggests that these new theories may
make possible the construction of large-$N$ models in small space-time
models, finally realizing the potential of the Eguchi-Kawai large-$N$
reduction \cite{Eguchi:1982nm,Kovtun:2007py,Bringoltz:2008av}.

When one or more directions are compact, the perturbative contribution
of the gauge fields to the effective potential favors the deconfined
phase. This leads to the well-known deconfinement transition in finite-temperature
gauge theories. Thus it is necessary to modify the gauge theory in
some way to obtain confinement with small compact directions. At present,
there are two methods known for achieving this. The first method directly
modifies the gauge action with terms non-local in the compact direction(s)
\cite{Myers:2007vc},
while the second adds adjoint fermions with periodic boundary conditions
in the compact direction(s) \cite{Unsal:2007vu}, which is our subject here.


The term adjoint QCD is often used to refer to vector gauge theories
based on the gauge group $SU(N)$ where, in addition to the gauge
fields, fermions in the adjoint representation are fundamental fields.
We will use $N_{f}$ to denote the number of flavors of adjoint Dirac
fermions. With this notation $N_{f}=1/2$ corresponds to $\mathcal{N}=1$
supersymmetry when the mass  of the fermion is taken to zero.
If $N_{f} \le 2$, the theory will be asymptotically free.
Here we will consider models on $R^{3}\times S^{1}$ that give rise
to a confining phase if the circumference $L$ of $S^{1}$ is sufficiently
small and the mass $m$ of the adjoint fermions is sufficiently light.

Confinement in $SU(N)$ gauge theories is associated with an
unbroken global center symmetry, which is $Z(N)$ for $SU(N)$.
The deconfined phase is associated with the breaking of this
$Z(N)$ symmetry, which occurs spontaneously at high temperatures
 \cite{Gross:1980br,Weiss:1980rj}.
The
order parameter for $Z(N)$ breaking in the compact direction is the
Polyakov loop $P$, which is the path-ordered exponential of the gauge
field in the compact direction\begin{equation}
P\left(\vec{x}\right)=\mathcal{P}\exp\left[i\int_{0}^{L}dx_{4}A_{4}\left(x\right)\right].\end{equation}
The trace of $P$ in a representation $R$ represents the insertion
of a heavy fermion in that representation into the system. Unbroken
$Z(N)$ symmetry implies $\langle Tr_{F}P\rangle=0$ in the confined
phase, and correspondingly $\left\langle Tr_{F}P\right\rangle \ne0$ holds
in the deconfined phase where $Z(N)$ symmetry is broken.
The character $Tr_{R}P$ of each irreducible
representation of $SU(N)$ transforms as $Tr_{R}P\rightarrow z^{k}Tr_{R}P$
under $P\rightarrow zP$ for some $k\in\left\{ 0,..,N-1\right\} $.
In the confined phase, the expectation value $\left\langle Tr_{R}P\right\rangle $
is $0$ for all representations that transform non-trivially under
$Z(N)$, \emph{i.e.}, have $k\neq 0$.
In the case where bosons have periodic boundary conditions and fermions
have antiperiodic boundary conditions in the compact direction, the
transfer matrix in that direction is positive-definite and the system
can be interpreted as being at temperature $T=\beta^{-1}=L^{-1}$.
In gauge models where all physical fields have zero $N$-ality, the
free energy $F$ for a single test fermion in the fundamental representation 
is related to the expectation value of Polyakov loop\begin{equation}
e^{-\beta F}=<Tr_{F}P>.\end{equation}

The restoration of $Z(N)$ symmetry for small $L$ and small $m$
when the adjoint fermions have periodic boundary conditions
is seen from the behavior of the one-loop effective potential. This
calculation is a variant of the calculation of the effective
potential at finite temperature \cite{Gross:1980br,Weiss:1980rj,Meisinger:2001fi}. 
The sum of the effective potential
for the fermions plus that of the gauge bosons gives\begin{equation}
V_{1-loop}\left(P,L,m,N_{f}\right)=\frac{1}{\pi^{2}L^{4}}\sum_{n=1}^{\infty}\frac{Tr_{A}P^{n}}{n^{2}}\left[2N_{f}L^{2}m^{2}K_{2}\left(nLm\right)-\frac{2}{n^{2}}\right].\end{equation}
Note that the first term in brackets, due to the fermions, is positive
for every value of $n$, while the second term, due to the gauge bosons,
is negative. In the limit $m\rightarrow0$, this becomes\begin{equation}
\mbox{\ensuremath{V_{1-loop}\approx\sum_{j,k=1}^{N}}(1-\ensuremath{\frac{1}{N}\delta_{jk}})\ensuremath{\frac{2\left(2N_{f}-1\right)}{\pi^{2}L^{4}}\left[\frac{\pi^{4}}{90}-\frac{1}{48\pi^{2}}\left(\phi_{j}-\phi_{k}\right)^{2}\left(\phi_{j}-\phi_{k}-2\pi\right)^{2}\right]}}\end{equation}
where we have written $P$ as $P_{jk}=\delta_{jk}e^{i\phi_{j}}$ in
a gauge where the background field $A_{4}$ is diagonal and 
independent of $x_4$.
If $mL$ is sufficiently small,
this effective potential has a global minimum when the Polyakov loop
eigenvalues are uniformly spaced around the unit circle. This is the
unique $Z(N)$-symmetric solution for $P$. On the other hand, if
$mL$ is sufficiently large, the fermion contribution to $V_{1-loop}$
is negligible compared to the gauge boson contribution, and the system
will be in the deconfined phase. Interestingly, there can be intermediate
phases between the confined and deconfined phases as $m$ is varied
for $N\ge3$  \cite{Myers:2007vc,Myers:2009df}. 
It is also interesting to note that the one-loop effective
potential identically vanishes when $N_{f}=1/2$ in the $m=0$ limit,
consistent with the supersymmetry of the model \cite{Davies:1999uw,Davies:2000nw}.

It is clearly crucial that the fermion mass $m$ be sufficiently small
in order for confinement to be restored at small $L$. Experience
with phenomenological models 
\cite{Gocksch:1984yk,Fukushima:2003fw}
suggests that in fact it is the constituent
mass which is relevant in determining the size of the fermionic contribution
to the effective potential for $P$. The potential importance of chiral
symmetry breaking and restoration is underscored by lattice results
for adjoint $SU(3)$ fermions at finite temperature, where the fermions
are antiperiodic in the Euclidean timelike direction. In this case,
there is a large separation between the deconfinement temperature
and the chiral symmetry restoration temperature, 
with $T_{c}/T_{d}\simeq7.8(2)$ \cite{Karsch:1998qj,Engels:2005te}.


In order to explore the interrelationship of confinement and chiral
symmetry breaking, we use a generalization of Nambu-Jona Lasinio models
known as Polyakov-Nambu-Jona Lasinio (PNJL) models  \cite{Fukushima:2003fw}. 
In NJL models,
a four-fermion interaction induces chiral symmetry breaking. There
has been a great deal of work on NJL models, both as phenomenological
models for hadrons and as effective theories of QCD 
\cite{Klevansky:1992qe,Hatsuda:1994pi}.
NJL models have been used to study hadronic physics at
finite temperature, but they include only chiral symmetry restoration,
and do not model deconfinement. This omission is rectified by the
PNJL models, which include both chiral restoration and deconfinement.
The earliest model of this type was derived from strong-coupling lattice
gauge theory \cite{Gocksch:1984yk}, but later work on continuum models have
proven to be extremely powerful in describing the finite-temperature
QCD phase transition \cite{Fukushima:2003fw}. In PNJL models, fermions with
NJL couplings move in a non-trivial Polyakov loop background, and
the effects of gluons at finite temperature is modeled in a semi-phenomenological
way. We will develop a model of this type for both fundamental
and adjoint fermions below.


These models will show why $mL$, with $m$ interpreted
as  a constituent mass, is the key parameter.
Because we expect
$m$ at worst to stay constant
as $L$ is decreased, the condition
for confinement that $mL$ is small 
can always be met at sufficiently
small $L$. It is not \emph{a priori} obvious what sufficiently small
means, given the persistence of chiral symmetry breaking in the deconfined
phase. 
Recent lattice simulations by Cossu and D'Elia \cite{Cossu:2009sq}
have confirmed the existence of the small-$L$ confined region in
$SU(3)$ lattice gauge theory with two flavors of adjoint fermions,
and we will focus on this case in our analysis.
Even if
the small-$L$ confined region exists and is accessible in lattice
simulations, it is not necessarily the same phase as found for large
$L$. Put slightly differently, we would like to know if the small-$L$
and large-$L$ confined regions are smoothly connected, and thus represent
the same phase. Our main result
will be a phase diagram for adjoint periodic QCD for all values of $L$, 
obtained using a PNJL model.
On the way to this goal, we will use as tests of our model
both standard QCD with fundamental fermions 
and adjoint QCD with the usual antiperiodic boundary
conditions for fermions. Our principal
tool will be the effective potential for the chiral symmetry order
parameter $\bar{\psi}\psi$ and the deconfinement order parameter
$P$. 

In section II  the fermionic contribution to the effective potential is derived,
and section III discusses the gluonic contribution. The complete effective potential
is tested with fundamental fermions in section IV. Sections V and VI discuss
adjoint fermions with antiperiodic and periodic boundary conditions, respectively.
A final section summarizes our conclusions.

\section{Fermionic contribution to effective potential}

We start our calculations by constructing the contributions to the
effective potential from both fundamental and adjoint fermions. 
NJL models use purely
fermionic interactions as a proxy for the gauge theory interactions
that give rise to chiral symmetry breaking. Typically, the relation
between the gauge theory and an associated NJL model is fixed by matching
important hadronic parameters such as $f_{\pi}$. In the case of NJL
models at finite temperature, and particularly PNJL models, it is
common to assume that the NJL model parameters are fixed by $T=0$
hadronic parameters and remain constant as $T$ is increased, at least
up to the deconfinement temperature. At substantially higher temperatures,
it is possible that the parameters of the NJL model might have to
be adjusted to reproduce properties of the gauge theory at those higher
temperatures. This possibility is particularly acute in adjoint QCD,
where the scale of chiral symmetry restoration is almost a factor
of ten higher than the deconfinement scale \cite{Karsch:1998qj,Engels:2005te}. 
Therefore,
it is desirable to consider the phase diagram of the appropriate PNJL
model over a range of parameters, holding open the possibility that
the PNJL parameters are varied with $\beta$ or $L$.

We take the fermionic part of the Lagrangian of our PNJL model to be 
\cite{Klevansky:1992qe,Hatsuda:1994pi,Fukushima:2003fw}
\begin{equation}
L_{F}=\bar{\psi}\left(i\gamma\cdot D-m_{0}\right)\psi+\frac{g_{S}}{2}\left[\left(\bar{\psi}\lambda^{a}\psi\right)^{2}+\left(\bar{\psi}i\gamma_{5}\lambda^{a}\psi\right)^{2}\right]+g_{D}\left[\det\bar{\psi}\left(1-\gamma_{5}\right)\psi+h.c.\right]\end{equation}
where $\psi$ is associated with $N_{f}$ flavors of Dirac fermions
in the fundamental or adjoint representation of the gauge group $SU(N)$.
The $\lambda^{a}$'s are the generators of the flavor symmetry group
$U(N_{f})$; $g_{S}$ represents the strength of the four-fermion
scalar-pseudoscalar coupling and $g_{D}$ fixes the strength of an
anomaly-induced term. For simplicity, we take the Lagrangian mass
matrix $m_{0}$ to be diagonal: $\left(m_{0}\right)_{jk}=m_{0j}\delta_{jk}$.
If $m_{0}$ and $g_{D}$ are taken to be zero, $L_{F}$ is invariant
under the global symmetry $U(N_{f})_{L}\times U(N_{f})_{R}$. With
a non-zero, flavor-independent mass $m_{0}$ and $g_{D} \neq 0$, the symmetry is
reduced to $SU(N_{f})_{V}\times U(1)_{V}$.
The covariant derivative $D_{\mu}$ couples the fermions to a background Polyakov
loop via the component of the gauge field in the compact direction.

We will use the identity\begin{equation}
\lambda_{ij}^{a}\lambda_{kl}^{a}=2\delta_{il}\delta_{jk}\end{equation}
for the generators of $U(N_{f})$ to rewrite the scalar-scalar four-fermion
coupling as \begin{equation}
g_{S}\left[\left(\bar{\psi_{i}}\psi_{j}\right)\left(\bar{\psi_{j}}\psi_{i}\right)\right].\end{equation}
The terms with $i=j$

\begin{equation}
g_{S}\sum_{i}\left(\bar{\psi_{i}}\psi_{i}\right)^{2}\end{equation}
 give rise to an interaction that does not mix flavors. These terms
are of order $N^{2}$, as opposed to the terms with $i\ne j$,
which are only of order $N$. Keeping only the terms with $i=j$
is the Hartree approximation, but the $i\ne j$ contribution to the
one-loop vacuum energy is canceled by the pseudoscalar interaction
in any case. We will make a similar approximation for the $g_{D}$
interaction.

The partition function associated with $L_{F}$ is\begin{equation}
Z_{F}=\int\left[d\bar{\psi}\right]\left[d\psi\right]e^{i\int d^{4}x\, L_{F}}\end{equation}
and depends implicitly on the background gauge field via the covariant
derivative. We introduce a set of auxiliary scalar fields, $\sigma_{j}$
and $\alpha_{j}$ for each flavor $j$, and write the partition function
as\begin{equation}
Z=\int\left[d\bar{\psi}\right]\left[d\psi\right]\left[d\sigma\right]\left[d\alpha\right]\exp\left[i\int d^{4}x\,\left(L_{F}+\sum_{j}\alpha_{j}\left(\sigma_{j}-\bar{\psi}_{j}\psi_{j}\right)\right)\right]\end{equation}
 where the $\alpha_{j}$'s provide functional $\delta$ functions
that set $\sigma_{j}$ equal to $\bar{\psi}_{j}\psi_{j}$. This allows
us to rewrite $Z$ in the form\begin{equation}
Z=\int\left[d\bar{\psi}\right]\left[d\psi\right]\left[d\sigma\right]\left[d\alpha\right]\exp\left\{ i\int d^{4}x\,\left[\sum_{j}\bar{\psi}_{j}\left(i\gamma\cdot D-m_{0j}-\alpha_{j}\right)\psi_{j}+\sum_{j}g_{S}\sigma_{j}^{2}+2g_{D}\prod_{j}\sigma_{j}+\sum_{j}\alpha_{j}\sigma_{j}\right]\right\} .\end{equation}
We proceed by integrating over the fermion fields to obtain\begin{equation}
Z=\int\left[d\sigma\right]\left[d\alpha\right]e^{iS_{eff}}\end{equation}
where the effective action $S_{eff}$ is given by\begin{equation}
S_{eff}=-i\sum_{j}\ln\left[\det\left(i\gamma\cdot D-m_{0j}-\alpha_{j}\right)\right]+\int d^{4}x\left[\sum_{j}g_{S}\sigma_{j}^{2}+2g_{D}\prod_{j}\sigma_{j}+\sum_{j}\alpha_{j}\sigma_{j}\right].\end{equation}
 We look for the stationary saddle points of the effective action
regarded as a function of the $\alpha_{j}$'s. The expression to be
minimized has the form\begin{equation}
\sum_{j}F_{j}(\alpha_{j})+\sum_{j}\alpha_{j}\sigma_{j}-V(\sigma)\end{equation}
whose saddle point solutions satisfy\begin{equation}
\frac{\partial F_{j}}{\partial\alpha_{j}}+\sigma_{j}=0\end{equation}
\begin{equation}
\alpha_{j}-\frac{\partial V}{\partial\sigma_{j}}=0\end{equation}
 which reduces to\begin{equation}
\left[\frac{\partial F_{j}}{\partial\alpha_{j}}\right]_{\alpha_{j}=\frac{\partial V}{\partial\sigma_{j}}}+\sigma_{j}=0.\end{equation}
 This in turn is equivalent to extremizing\begin{equation}
\sum_{j}F_{j}\left(\frac{\partial V}{\partial\sigma_{j}}\right)+\sum_{j}\sigma_{j}\frac{\partial V}{\partial\sigma_{j}}-V\end{equation}
In this particular case, $S_{eff}$ is reduced to\begin{eqnarray*}
S_{eff} & = & -i\sum_{j}\ln\left[\det\left(i\gamma\cdot D-m_{0j}+2g_{S}\sigma_{j}+2g_{D}\prod_{k\ne j}\sigma_{k}\right)\right]\\
 &  & +\int d^{4}x\left[-\sum_{j}g_{S}\sigma_{j}^{2}-2g_{D}\left(N_{f}-1\right)\prod_{j}\sigma_{j}\right].\end{eqnarray*}
We see immediately that each flavor behaves as if it has a mass given
by $m_{j}=m_{0j}-2g_{S}\sigma_{j}-2g_{D}\prod_{k\ne j}\sigma_{k}$.
The boundary conditions in the compact direction
enter the effective potential through the fermion
determinant, so it
is through the mass $m_{j}$, the constituent mass, that the Polyakov
loop and the chiral order parameters $\sigma_{j}$ are coupled.


It is generally convenient to use the language of finite temperature to describe
both the case of  finite temperature, $\beta^{-1}=T>0$, with antiperiodic boundary conditions,
and the case of a periodic spatial direction, $L<\infty$.
The fermionic contribution to the effective potential has both a zero-temperature
and a finite-temperature contribution. The zero-temperature part consists
of a potential term, given by
\begin{equation}
\sum_{j}g_{S}\sigma_{j}^{2}+2g_{D}\left(N_{f}-1\right)\prod_{j}\sigma_{j}
\end{equation}
as well as a contribution from the fermion functional determinant.
This is formally given by\begin{equation}
-2d_{R}\sum_{j=1}^{N_{f}}\int\frac{d^{3}k}{(2\pi)^{3}}\omega_{k}^{\left(j\right)}.\end{equation}
where $\omega_{k}^{\left(j\right)}=\sqrt{k^{2}+m_{j}^{2}}$, and the
constant $d_{R}$ is the dimensionality of the color representation,
$N$ for the fundamental and $N^{2}-1$ for the adjoint. This contribution,
representing a sum of one-loop diagrams, is divergent, and requires
regularization. Although many schemes have been used \cite{Klevansky:1992qe}, a non-covariant
three-dimensional cutoff has been used most often. With this regularization,
the zero-temperature part of the fermion determinant is given by\begin{equation}
-2d_{R}\sum_{j=1}^{N_{f}}\frac{2\text{\ensuremath{\Lambda}}\sqrt{\text{\ensuremath{\Lambda}}^{2}+m_{j}^{2}}\left(2\text{\ensuremath{\Lambda}}^{2}+m_{j}^{2}\right)+m_{j}^{4}\left[\log\left(m_{j}^{2}\right)-2\log\left(\Lambda+\sqrt{\Lambda^{2}+m_{j}^{2}}\right)\right]}{32\pi^{2}}\end{equation}
where $\Lambda$ is the momentum space cutoff. 
Thus the total $T=0$ fermionic contribution to the effective potential
is
\begin{eqnarray}
V_{F0}\left(m,m_{0,}\right)&=&
\sum_{j}g_{S}\sigma_{j}^{2}+2g_{D}\left(N_{f}-1\right)\prod_{j}\sigma_{j}\nonumber\\
&&-2d_{R}\sum_{j=1}^{N_{f}}\frac{2\text{\ensuremath{\Lambda}}\sqrt{\text{\ensuremath{\Lambda}}^{2}+m_{j}^{2}}\left(2\text{\ensuremath{\Lambda}}^{2}+m_{j}^{2}\right)+m_{j}^{4}\left[\log\left(m_{j}^{2}\right)-2\log\left(\Lambda+\sqrt{\Lambda^{2}+m_{j}^{2}}\right)\right]}{32\pi^{2}}
\end{eqnarray}


In PNJL models, the finite-temperature contribution from the fermion
determinant depends on the background Polyakov loop. It is convenient
to work in a gauge where the temporal component of the background
gauge field, $A_{4}(\vec{x},\, t)$, is constant and diagonal. The
covariant derivative then becomes $\gamma\cdot D=\gamma\cdot\partial-i\gamma^{4}A_{4}$.
The one-loop free energy of fermions in a representation $R$ of $SU(N)$
gauge theory with zero chemical potential can be written as
\begin{equation}
V_{FL}\left(P,m\right)=-2\sum_{j}Tr_{R}[\frac{1}{L}\int\frac{d^{3}k}{(2\pi)^{3}}ln(1\mp Pe^{-L\omega_{k}^{\left(j\right)}})+h.c.]
\end{equation}
where the minus sign is used for periodic boundary conditions and
plus for antiperiodic. In the latter case, $L$ is replaced by $\beta$. 
The potential term $V_{FT}$ has a series expansion
in terms of modified Bessel functions
\begin{equation}
V_{FL}\left(P,m\right)=\sum_{j}\frac{2m_{j}^{2}}{\pi^{2}L^{2}}\sum_{n=1}^{\infty}\frac{(\pm 1)^{n}Tr_{R}P^{n}}{n^{2}}K_{2}\left(nL m_{j}\right)\end{equation}
which is rapidly convergent for all values of the mass
\cite{Meisinger:2001fi}.


Although NJL models display chiral symmetry breaking and are plausibly
related to QCD-like theories, they do not capture exactly all the
features of the gauge theory to which they are related. There are
many ways of associating a given NJL model with a gauge theory, and
there is freedom in choosing both regularization scheme and parameters
of the two theories to match. Typically, the hadronic parameters used
represent hadronic physics at zero temperature, and those parameters
are held fixed up to the chiral transition and perhaps beyond. 
For the case of $N=3$ models with two and three flavors in the fundamental
representation, appropriate choices of $m_{0},$ $g_S$ and $\Lambda$
can capably model many features of hadron physics.
In what follows, we will take $N_{f}=2$, and take the masses $m_{0j}$
to be equal to a common mass which we also write as $m_{0}$. In this
case, the contribution to $S_{eff}$ from $g_{S}$ and $g_{D}$ has
the same form. It is convenient to take $g_{D}=0$, and also to write
the common constituent mass as $m=m_{0}-2g_{S}\sigma$ 
\cite{Klevansky:1992qe,Hatsuda:1994pi}.
In PNJL models,
additional information about confinement at non-zero temperature
must be added. As will be discussed below, we will use as input the
deconfinement temperature $T_{d}$ for the pure gauge theory.

At non-zero temperature, the observables of a given gauge theory will
evolve with $T$ according to the finite-temperature renormalization
group, but the parameters of an associated NJL or PNJL model do not
naturally evolve in a related way. If we follow the behavior of both
the gauge theory and an associated PNJL model over a large scale of
temperatures, it may be necessary or desirable to consider the parameters
of the PNJL model as varying with $T$. For example, Unsal has developed
a comprehensive scenario for periodic adjoint fermions with $m_{0}=0$
in which chiral symmetry is spontaneously broken by monopole-induced
anomaly terms \cite{Unsal:2007vu,Unsal:2007jx}. 
If this interaction is sufficiently weak, it may
not induce chiral symmetry breaking, raising the possibility of confinement
without chiral symmetry breaking. There is also the possibility of
modifying the strength of chiral symmetry breaking by adding additional
couplings compatible with all symmetries have been added. The easiest
construction of such an extended model might be obtained by adding
to a gauge theory a non-renormalizable four-fermion coupling of exactly
the same form as in an NJL model, but with opposite sign for the
coupling: \begin{equation}
L_{gauge}\rightarrow L_{gauge}+\frac{\delta g_{S}}{2}\left[\left(\bar{\psi}\lambda^{a}\psi\right)^{2}+\left(\bar{\psi}i\gamma_{5}\lambda^{a}\psi\right)^{2}\right].\end{equation}
If we imagine that the gauge theory is associated with an NJL model
$L_{NJL}$ with a certain value of $g_{S}$, then the additional term
shifts the NJL Lagrangian as\begin{equation}
L_{NJL}\rightarrow L_{NJL}+\frac{\delta g'_{S}}{2}\left[\left(\bar{\psi}\lambda^{a}\psi\right)^{2}+\left(\bar{\psi}i\gamma_{5}\lambda^{a}\psi\right)^{2}\right].\end{equation}
where one expects $\delta g'_{S}=\delta g_{S}$ for a sufficiently
small perturbation. In principle, a lattice version of the extended
gauge theory could be simulated, and the results combined with analytical
results for the NJL model to determine $\delta g'_{S}$ as a function
of $\delta g_{S}$. In fact, lattice gauge theories with additional
non-renormalizable four-fermion terms added have already been used
in the study of finite temperature gauge theories 
\cite{Brower:1994sw,Brower:1995vf,Kogut:1998rg}.
For the two-flavor case we
discuss, it is useful to treat all parameters of the model as potentially
varying.

\section{Gluonic contribution to effective potential}

\begin{figure}
\includegraphics[width=5in]{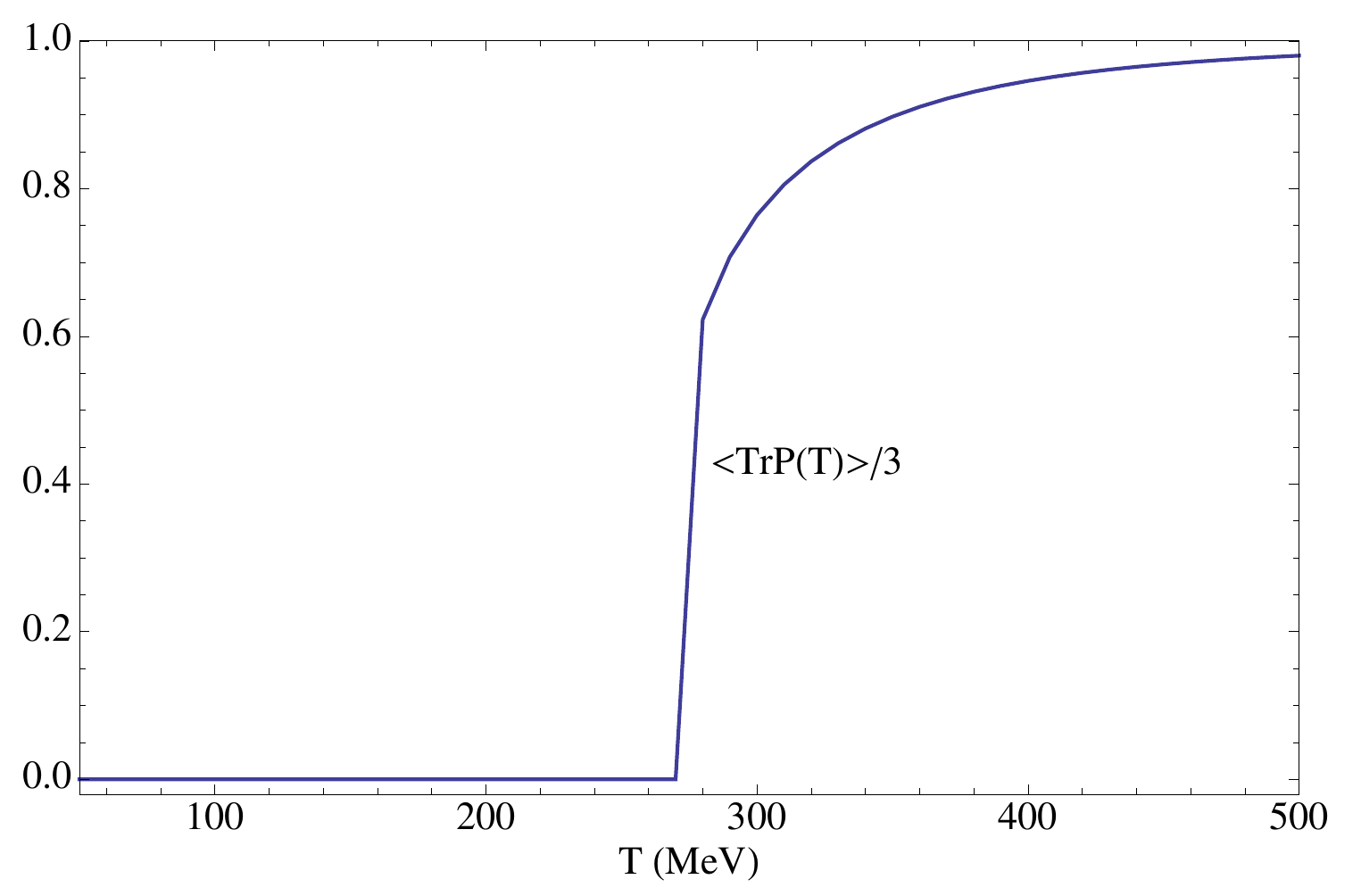}
\caption{The normalized order parameter $\langle Tr_F P \rangle/3$ for pure $SU(3)$ gauge theory as a function of temperature. The deconfinement transition is first-order,  and occurs at $T_d = 270\,MeV$.}
\label{fig:TvsTrP_Gluonic}
\end{figure}
For gauge bosons, the one-loop finite temperature free energy in a
background Polyakov loop is given by an expression similar to the
one for fermions. The boundary conditions for the gauge bosons
are periodic in all cases considered here, so $L$ and $\beta$
may be used equivalently in the gluonic sector.
We have\begin{equation}
V_{g-1\,loop}\left(P\right)=2\, Tr_{A}[\frac{1}{L}\int\frac{d^{3}k}{(2\pi)^{3}}ln(1-Pe^{-L\Omega_{k}})]\end{equation}
where we have inserted a mass parameter in $\Omega_{k}=\sqrt{k^{2}+M^{2}}$
for purely phenomenological reasons explained below. If we take the
zero mass limit of the series expansion, we get the standard expression
for the one-loop gauge boson free energy \cite{Gross:1980br,Weiss:1980rj}
\begin{equation}
-\lim_{M\rightarrow0}\frac{M^{2}}{\pi^{2}L^{2}}\sum_{n=1}^{\infty}\frac{Tr_{A}P^{n}}{n^{2}}K_{2}\left(nLM\right)=-\frac{2}{\pi^{2}L^{4}}\sum_{n=1}^{\infty}\frac{Tr_{A}P^{n}}{n^{4}}.\end{equation}
This infinite series may be summed exactly, giving an expression for
$V_{g-1\,loop}$ proportional to $L^{-4}$ involving the fourth Bernoulli
polynomial. If we also retain the next-order term in a high-temperature
expansion, proportional to $M^{2}/L^{2}$, we obtain a useful phenomenological
model for the deconfinement transition in pure gauge theories \cite{Meisinger:2001cq}.
The potential takes the form\begin{equation}
V_{g}\left(P\right)=-\frac{2}{\pi^{2}L^{4}}\sum_{n=1}^{\infty}\frac{Tr_{A}P^{n}}{n^{4}}+\frac{M^{2}}{2\pi^{2}L^{2}}\sum_{n=1}^{\infty}\frac{Tr_{A}P^{n}}{n^{2}}\end{equation}
We stress that the mass parameter $M$ should not be interpreted
as a gauge boson mass, nor do we limit ourselves to $ML\ll1$. The
crucial feature of this potential is that for sufficiently large values
of the dimensionless parameter $ML$, the potential leads to a $Z(N)$-symmetric,
confining minimum for $P$ \cite{Myers:2009df,Meisinger:2009ne}. 
On the other hand, for small values
of $ML$, the pure gauge theory will be in the deconfined phase. It
will be important later that $V_g$ is a good
representation of the gauge boson contribution for high temperatures;
in other PNJL models, the gauge boson contribution has sometimes
been chosen so as to be valid over a more narrow range of temperatures.
Both of the infinite sums
can be carried out exactly, giving a closed form for $V_{g}$ as a
function of the angles $\phi_{j}$ in terms of the fourth and second
Bernoulli polynomials \cite{Meisinger:2001cq}.


In the Polyakov gauge, the Polyakov loop in the fundamental representation
of $SU(3)$ can be written as $ $$P_{jk}=exp(i\phi_{j})\delta_{jk}$
with two independent angles. With the use of $Z(3)$ symmetry,
it is sufficient to consider the case where
$\left\langle Tr_{F}P\right\rangle $ is real.
Thus we consider only
diagonal, special-unitary matrices with real trace, which may be parametrized
by taking $\phi_{1}=\phi$, $\phi_{2}=-\phi$, and $\phi_{3}=0$,
or $P=diag\left[e^{i\phi},e^{-i\phi},1\right]$ with $0\le\phi\le\pi$.
The unique set of $Z(3)$-invariant eigenvalues are obtained for $\phi=2\pi/3$.
For $SU(3)$, $V_{g}$ takes the form:
\begin{equation}
V_{g}\left(P\right)=\left(\frac{3\phi^{2}}{2\pi^{2}}-\frac{2\phi}{\pi}+\frac{2}{3}\right)\frac{M^{2}}{L^{2}}+\frac{1}{L^{4}}\left(\frac{135\phi^{4}-300\pi\phi^{3}+180\pi^{2}\phi^{2}-16\pi^{4}}{90\pi^{2}}\right)\end{equation}
We will set the mass scale $M$ by requiring that $V_{g}$ yields the correct
deconfinement temperature for the pure gauge theory,
with a value of  $T_d\approx270\, MeV$. This gives $M=596\, MeV$ \cite{Meisinger:2001cq}.
The pressure $p$ is given by the value of $-V_{g}$ at the minimum
of the potential. The behavior obtained for the pressure, the energy
density $\epsilon$, and the so-called interaction measure $\Delta\equiv\left(\epsilon-3p\right)/T^{4}$
are all consistent with lattice simulations for $T>T_d$. 
As shown in figure \ref{fig:TvsTrP_Gluonic}, 
the order parameter $TrP$ for the deconfinement transition jumps
at $T_d$, indicating a first-order deconfinement transition for $SU(3)$.

\section{Fundamental Fermions}

\begin{figure}
\includegraphics[width=5in]{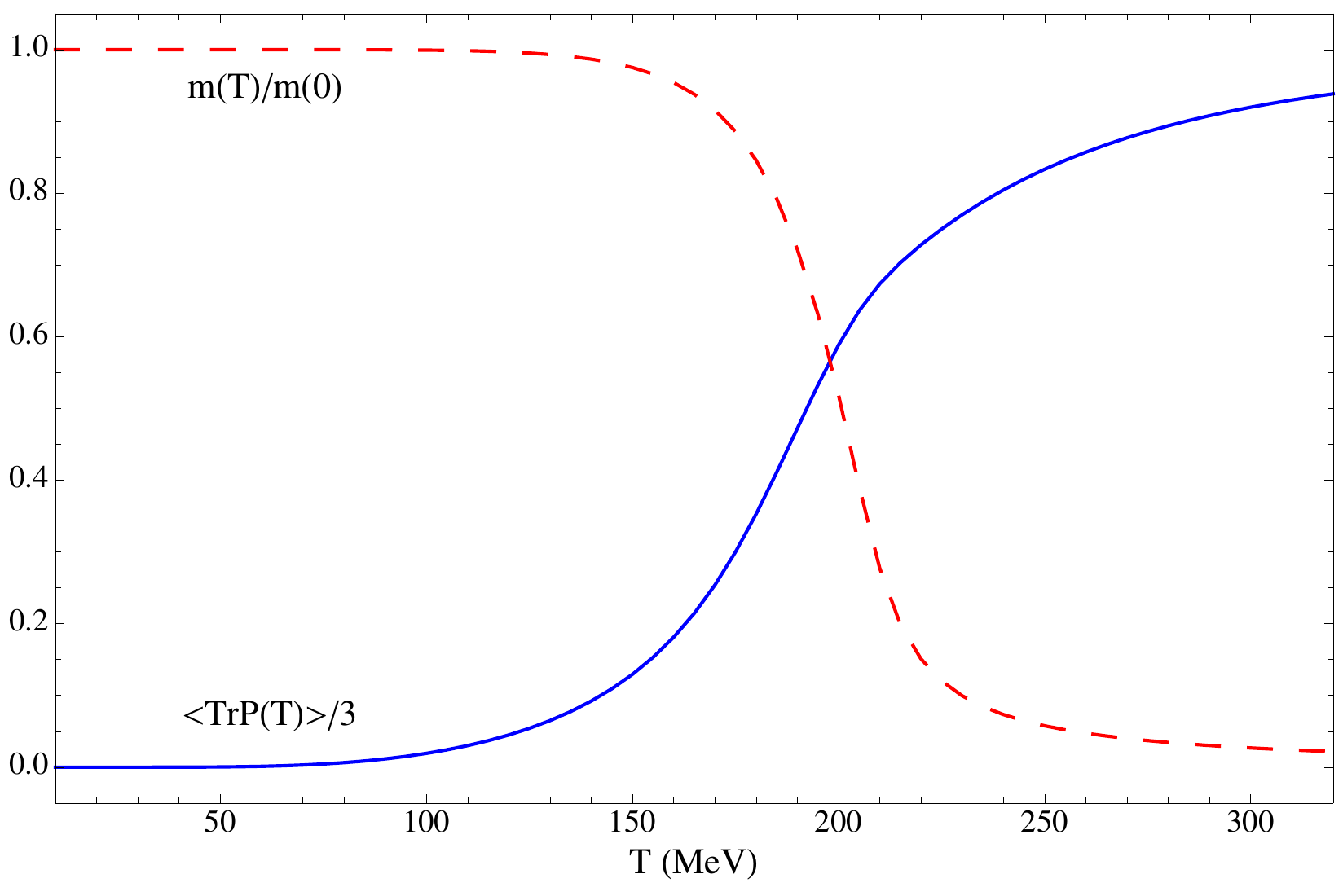}
\caption{The constituent mass $m$ and $\langle Tr_F P \rangle$ for two-flavor
QCD with fundamental representation fermions with antiperiodic boundary conditions
as a function of temperature. 
The order parameters are normalized by dividing by their values at $T=0$ and $T=\infty$, respectively.}
\label{fig:TvsOPs_FundABC}
\end{figure}
As a test of all the components of the effective potential we have
assembled, we consider the case of two flavors of fundamental fermions
at finite temperature. The fermions obey antiperiodic boundary conditions,
so we identify the compact circumference $L$ with the inverse
temperature $\beta=1/T$. A very common choice of zero-temperature
parameters for two degenerate light flavors is $m_{0}=5.5MeV,$ $\Lambda=631.4\,MeV,$
and $g_{S}=2\times5.496GeV^{-2}$
\cite{Hatsuda:1994pi,Fukushima:2003fw}.
In figure \ref{fig:TvsOPs_FundABC}, we show the 
expectation value of the Polyakov loop in the fundamental representation
$Tr_F P$ and the constituent mass $m$ as functions of the
temperature. The behavior in the crossover region is very similar
to the results of Fukushima \cite{Fukushima:2003fw}, and
shows the explanatory power of PNJL models. The constituent mass $m$
is heavy at low temperatures, due to chiral symmetry breaking. The
larger the constituent mass, the smaller the $Z(3)$ breaking effect
of the fermions. On the other hand, a small value for $\left\langle Tr_{F}P\right\rangle $
reduces the effectiveness of finite-temperature effects in restoring chiral symmetry.
These synergistic effects combine in the case of fundamental representation
fermions to give a single crossover temperature at which both order
parameters are changing rapidly, in agreement with lattice simulations.

\section{Adjoint Fermions with  Antiperiodic Boundary Conditions}

Adjoint $SU(3)$ fermions at finite temperature show a completely
different behavior in lattice simulations from fundamental fermions. 
Because the adjoint fermions
respect the $Z(3)$ center symmetry, there is a true deconfinement
transition where $Z(3)$ spontaneously breaks.
Lattice simulations have shown that chiral symmetry is
restored at a substantially higher temperature than
the deconfinement temperature \cite{Karsch:1998qj,Engels:2005te}.
Unlike the case of fundamental fermions, there are no
comprehensive $T=0$ lattice results for  adjoint $SU(3)$ fermions, no
established parameter sets for NJL models and,
of course, no experimental results to provide guidance.
We will again consider the case $N_{f}=2$ with degenerate
fermion masses. The $T=0$ parameters
needed are $g_{S}$ and $\Lambda$. The mass parameter $M$
is again fixed by the deconfinement transition in the pure gauge theory,
and $g_{D}$ is again absorbed into $g_S$. 
Rather than work directly with
$g_{S}$, we will consider the dimensionless coupling $\kappa=g_{S}\Lambda^{2}$.
A given ratio of $m(T=0)/\Lambda$ determines
the value of $\kappa$, and vice versa.
The value of $\Lambda$ is determined by the requirement that
$T_{c}/T_{d}$ is near $7.8$ \cite{Engels:2005te}. This in turn determines the value of the constituent
mass for all $T$.

\begin{figure}
\includegraphics[width=5in]{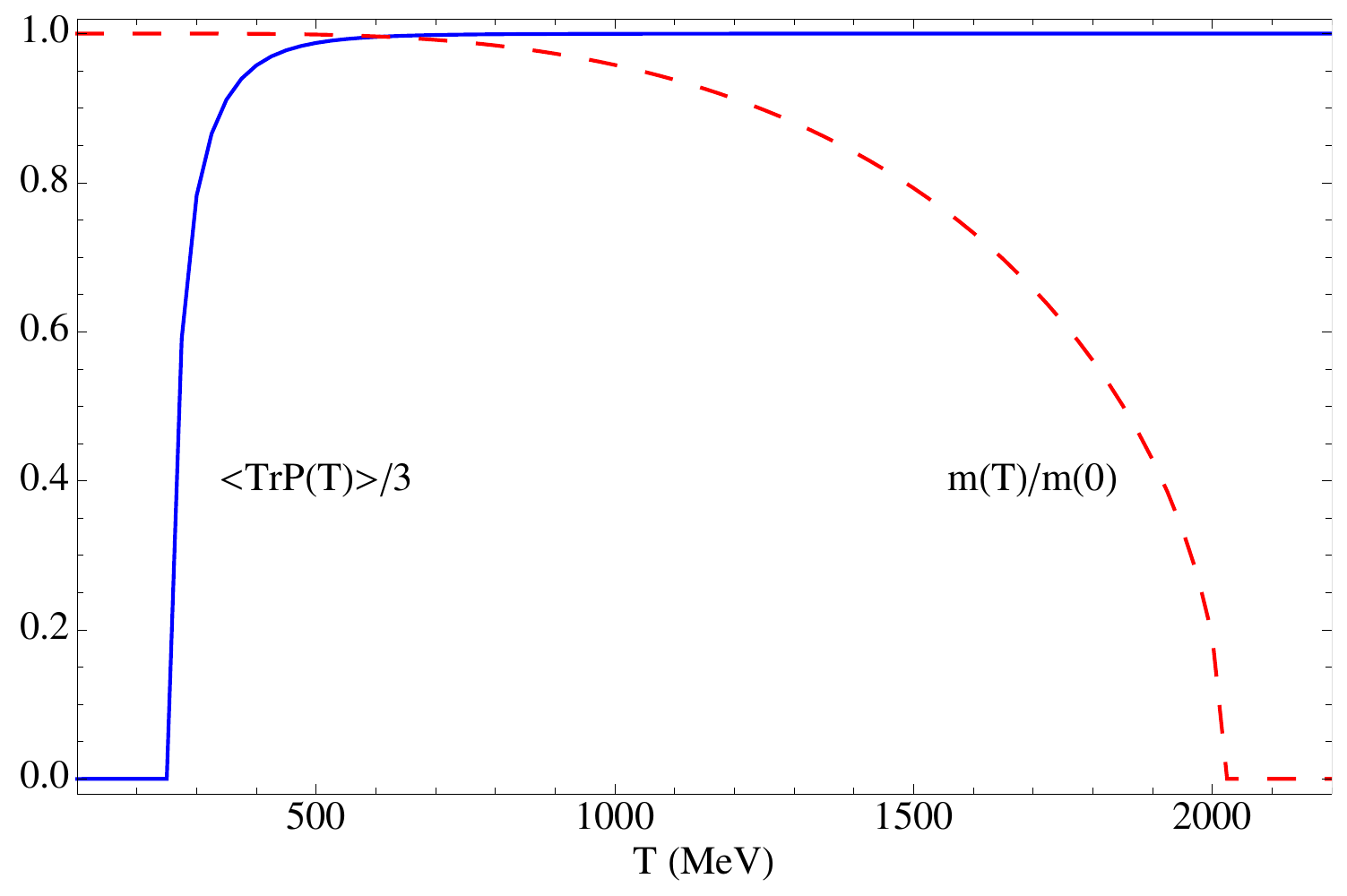}
\caption{The constituent mass $m$ and $\langle Tr_F P \rangle$ for two-flavor
QCD with adjoint representation fermions with antiperiodic boundary conditions
as a function of temperature for one choice of $\kappa$ and $\Lambda$. 
The order parameters are normalized by dividing by their values at $T=0$ and $T=\infty$, respectively.}
\label{fig:TvsOPs_AdjABC}
\end{figure}

We expect that the constituent mass
$m(T=0)$ must be substantially larger
than the corresponding value for
fundamental fermions, because a large constituent mass
is necessary to delay the onset of chiral symmetry restoration.
On the other hand, a large constituent mass at the deconfining
transitions would be expected to lead to
a relatively small change in the deconfining temperature.
The ratio $m(T=0)/\Lambda$
should be less than one in
order for the cutoff theory to be meaningful.
In the case of fundamental fermions, this ratio
is relatively large, on the order of $0.5$.
We have generally found that for adjoint fermions
a larger ratio of $m(T=0)/\Lambda$
with $T_{c}/T_{d}$ fixed
implies a larger value of $m(T=0)$.
We will work with the representative
case of $m_{0}=0$ 
and $m(T=0)/\Lambda=0.1$. 
This gives $\Lambda=23.22\,GeV$ and thus 
$m(T=0)=2.322\,GeV$, with $\kappa=1.2653$. 
For comparison, the critical value of $\kappa$, $\kappa_c$,
below which $m(T=0)=0$, is $\pi^2/8\simeq 1.234$.
In figure \ref{fig:TvsOPs_AdjABC},
we show the constituent mass $m$ and 
Polyakov expectation value $\langle Tr_F P \rangle$ 
as a function of temperature,
normalized by dividing by their values at $T=0$ and $T=\infty$, respectively.
We see that the deconfinement temperature $T_d$
is very close to its value in the pure gauge theory,
due to the large adjoint fermion constituent mass.
The transition is first order.
The constituent mass $m$ has a slow decline
to a second-order transition at a substantially
higher temperature, as indicated by lattice simulations 
\cite{Karsch:1998qj,Engels:2005te}.

\section{Adjoint Fermions with Periodic Boundary Conditions}

Having established that our PNJL model can provide a useful representation
for the finite-temperature behavior of $SU(3)$ with both fundamental
and adjoint fermions, we turn now to our main interest, adjoint fermions
on $R^{3}\times S^{1}$ with periodic boundary conditions for the fermions
in the compact direction. The sole change in the total effective potential
from the finite-temperature case lies in the contribution of the adjoint
fermions, given by $V_{FL}$. This single change leads to an unusual
and unexpected phase structure as $L$ is 
decreased \cite{Myers:2007vc,Myers:2009df}.

The effective potential can be written as
\begin{eqnarray}
V_{eff}\left(m,P\right) & = & V_{F0}\left(m\right)+V_{FL}\left(m,P\right)+V_{g}\left(P\right)\nonumber\\
 & = & V_{F0}\left(m\right)+\sum_{n=1}^{\infty}\left[\sum_{j}\frac{2m_{j}^{2}L^{2}}{\pi^{2}n^{2}}K_{2}\left(nLm_{j}\right)+\frac{M^{2}L^{2}}{2\pi^{2}n^{2}}-\frac{2}{\pi^{2}n^{4}}\right]\frac{Tr_{A}P^{n}}{L^{4}}
 \end{eqnarray}
 which is general for the case of adjoint fermions with periodic boundary
conditions in $SU(N)$. 
This equation contains the basic physics of the phase diagram.
The contributions of the first two terms in square
brackets are always positive, while the final term is always negative.
Consider first the $n=1$ term\begin{equation}
\left[\sum_{j}\frac{2m_{j}^{2}L^{2}}{\pi^{2}}K_{2}\left(Lm_{j}\right)+\frac{M^{2}L^{2}}{2\pi^{2}}-\frac{2}{\pi^{2}}\right]\frac{Tr_{A}P}{L^{4}}\end{equation}
 which is generally the largest term in the series. If either $ML$
is sufficiently large, or some combination of $m_{j}L$'s is sufficiently
small, the overall sign of the expression in square brackets will
be positive. The $n=1$ term will then favor the minimization of $Tr_{A}P$,
giving $Tr_{A}P=-1$ and therefore $Tr_{F}P=0$. The first case, with
$L$ sufficiently large, can be identified with the usual low-temperature
confined phase of the pure gauge theory. The second case, however,
is capable of producing a confined phase at small $L$, provided the
constituent mass $m$ is sufficiently small. This will certainly occur
if chiral symmetry is restored, but it can also occur if $L$ is increased
even if $m$ is essentially constant. On the other hand, if the sign
of the coefficient of $Tr_{A}P$ in $V_{eff}$ is negative, the $n=1$
term will favor maximizing $Tr_{A}P$, which occurs in the deconfined
phase.
This simple picture is sufficient for $SU(2)$. The actual phase structure,
as predicted theoretically and found in lattice simulations, 
is more complicated when $N\ge3$ 
 \cite{Myers:2007vc,Myers:2009df,Cossu:2009sq}.
In the case of $SU(3)$,
there is an additional phase, the skewed phase, in which $Tr_{F}P$
is non-zero but negative. As $N$ increases, the number of possible
phases increases as well, giving rise to a rich phase structure
for small $L$ \cite{Myers:2009df}.

\begin{figure}
\includegraphics[width=5in]{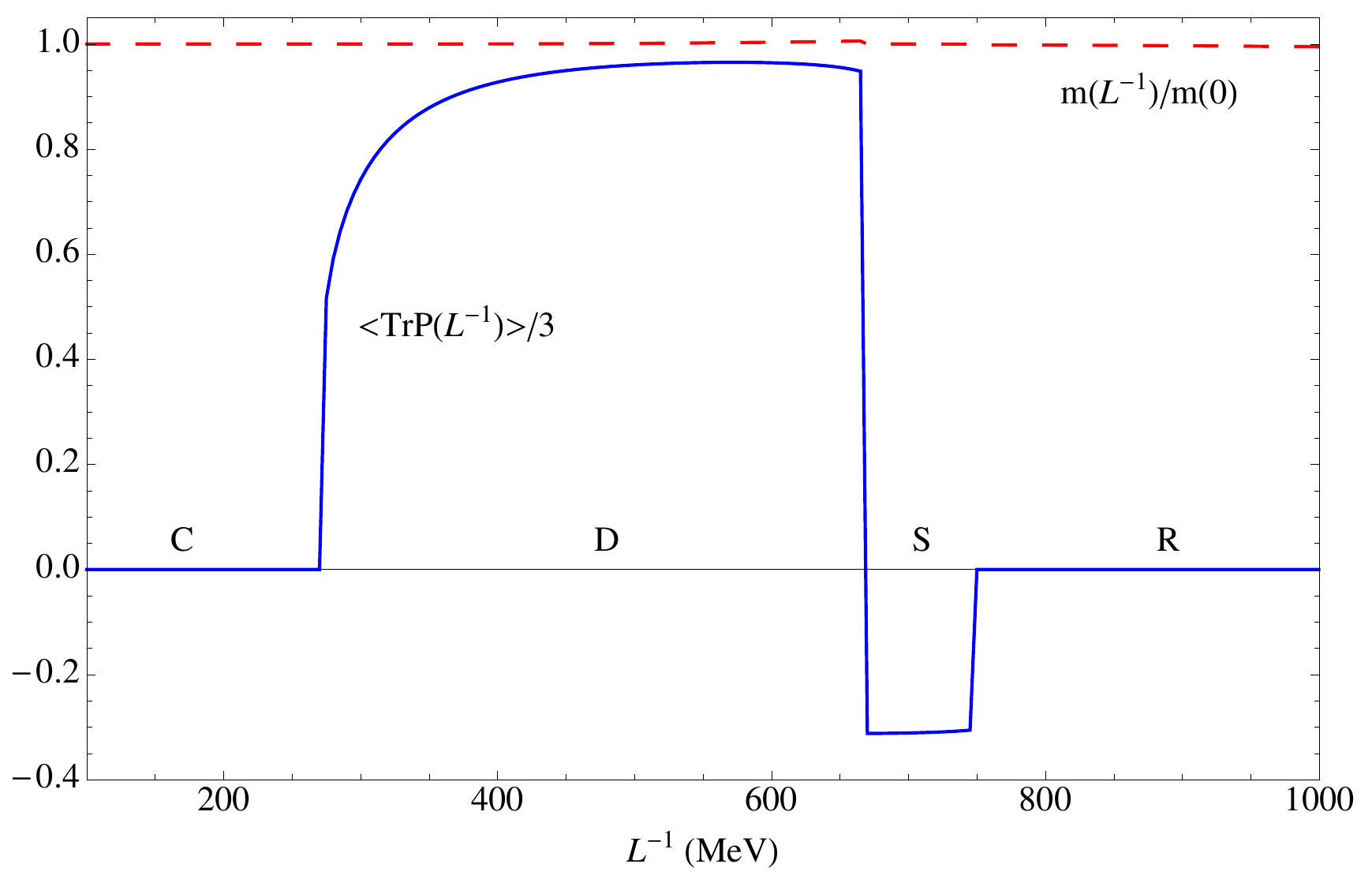}
\caption{The constituent mass $m$ and $\langle Tr_F P \rangle$ for two-flavor
QCD with adjoint representation fermions with periodic boundary conditions
as a function of $L^{-1}$ for one choice of $\kappa$ and $\Lambda$,
with $m_0 = 0$. 
The order parameters are normalized by dividing by their values at 
$L=\infty$ and $L=0$, respectively.
C, D, S and R refer to the confined phase, deconfined phase, skewed phase
and reconfined phase respectively.}
\label{fig:TvsOPs_AdjPBC}
\end{figure}

We consider the behavior of $m$ and $Tr_{F}P$ with periodic
fermions using the same parameters we used for the antiperiodic case.
Figure \ref{fig:TvsOPs_AdjPBC} shows the behavior of $m$ 
and $\langle Tr_F P\rangle$ as a function of $L^{-1}$ for
the $m(L=\infty)/\Lambda=0.1$ parameter set,
with $m_0=0$.
We see that chiral symmetry breaking
persists at $L^{-1}=10\, GeV$, which is much higher than the
chiral restoration temperature for antiperiodic fermions. 
The constituent mass
$m$ does fall eventually as $L^{-1}$ increases, and
chiral symmetry is ultimately restored,
but at a temperature on the order of $\Lambda$.
In figure
\ref{fig:TvsOPs_AdjPBC}, $Tr_{F}P$ shows
three distinct phase transitions as
a function of $L^{-1}$. As $L^{-1}$ increases, the confined phase
gives way to the deconfined phase in a first-order phase transition.
Because the constituent mass of the fermions is large, the critical
value of $L^{-1}$ for this transition is approximately equal to $T_d$.
As $L^{-1}$ increases, there are two more first-order
transitions, from the deconfined phase to the skewed phase, and then
from the skewed phase to a small-$L$ confined phase we describe 
as reconfined.

\begin{figure}
\includegraphics[width=5in]{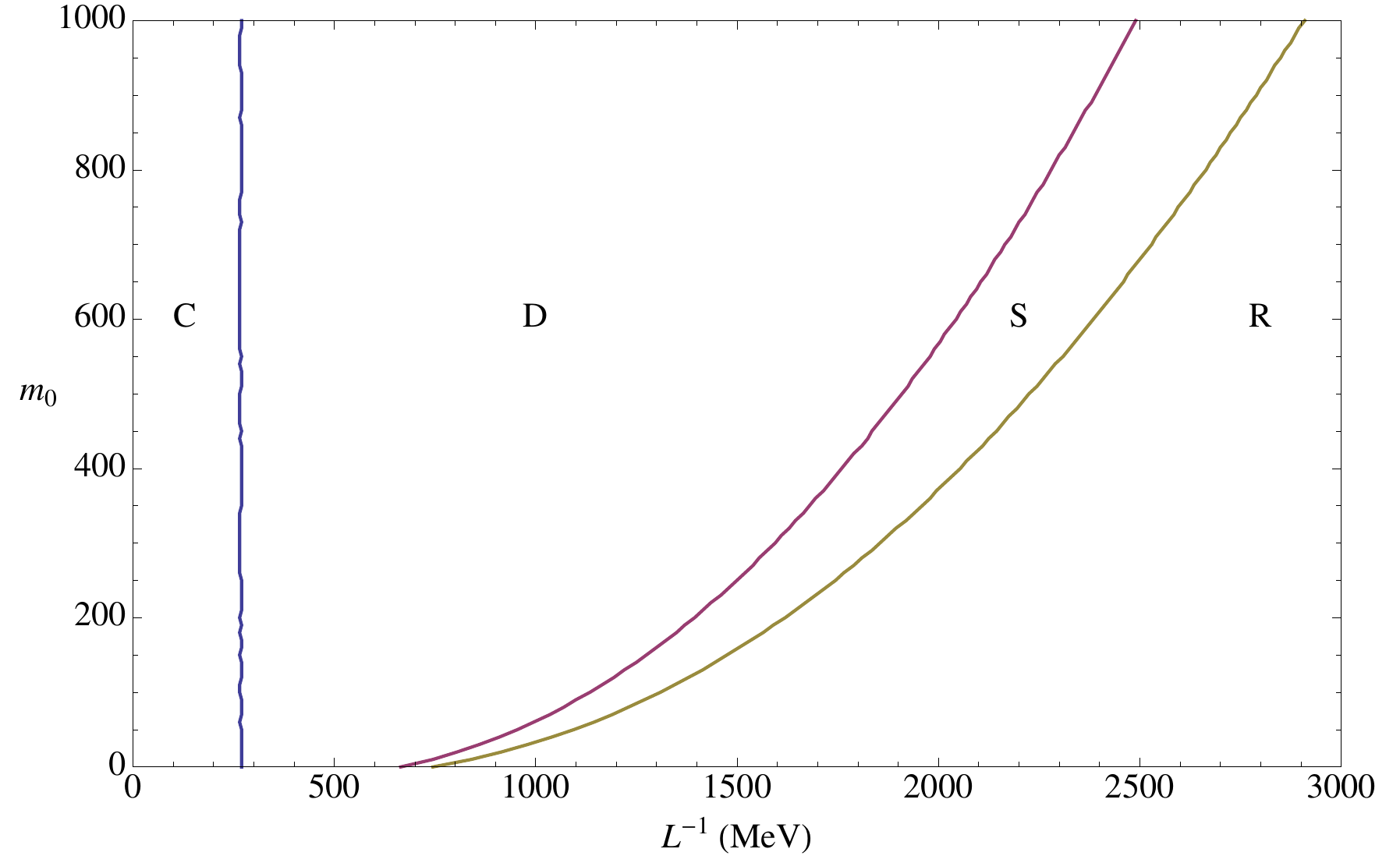}
\caption{The phase diagram for two-flavor
QCD with adjoint representation fermions with periodic boundary conditions
as a function of $L^{-1}$ and Lagrangian mass $m_0$
for one choice of $\kappa$ and $\Lambda$.
C, D, S and R refer to the confined phase, deconfined phase, skewed phase
and reconfined phase respectively. The Lagrangian mass is
measured in MeV.}
\label{fig:TvsM0_AdjPBC}
\end{figure}

The ordering of the phases seen in the behavior of $Tr_{F}P$ for
$m_{0}=0$ persists as $m_{0}$ is increased. Figure
\ref{fig:TvsM0_AdjPBC} shows 
that with this parameter set
the value of $L^{-1}$ for the confinement-deconfinement
transition stays essentially at $T_d$ as $m_{0}$ increases. The
only significant change in the phase diagram is the smooth growth
of the extent of the skewed phase as $m_{0}$ increases. Naively,
one might expect that the skewed and reconfined phases would disappear
as $m_{0}\rightarrow\infty$. However, at any fixed $m_{0}$, there
will be a sufficiently large value of $L^{-1}$ such that the fermionic
term in the effective potential overwhelms the gauge boson contribution,
and the skewed and reconfined phases do not disappear, but move to
very large values of $L^{-1}$. On the other hand, taking the limit
$m_{0}\rightarrow\infty$ at fixed $L^{-1}$ greater than $T_d$
always yields the deconfined phase.

\begin{figure}
\includegraphics[width=5in]{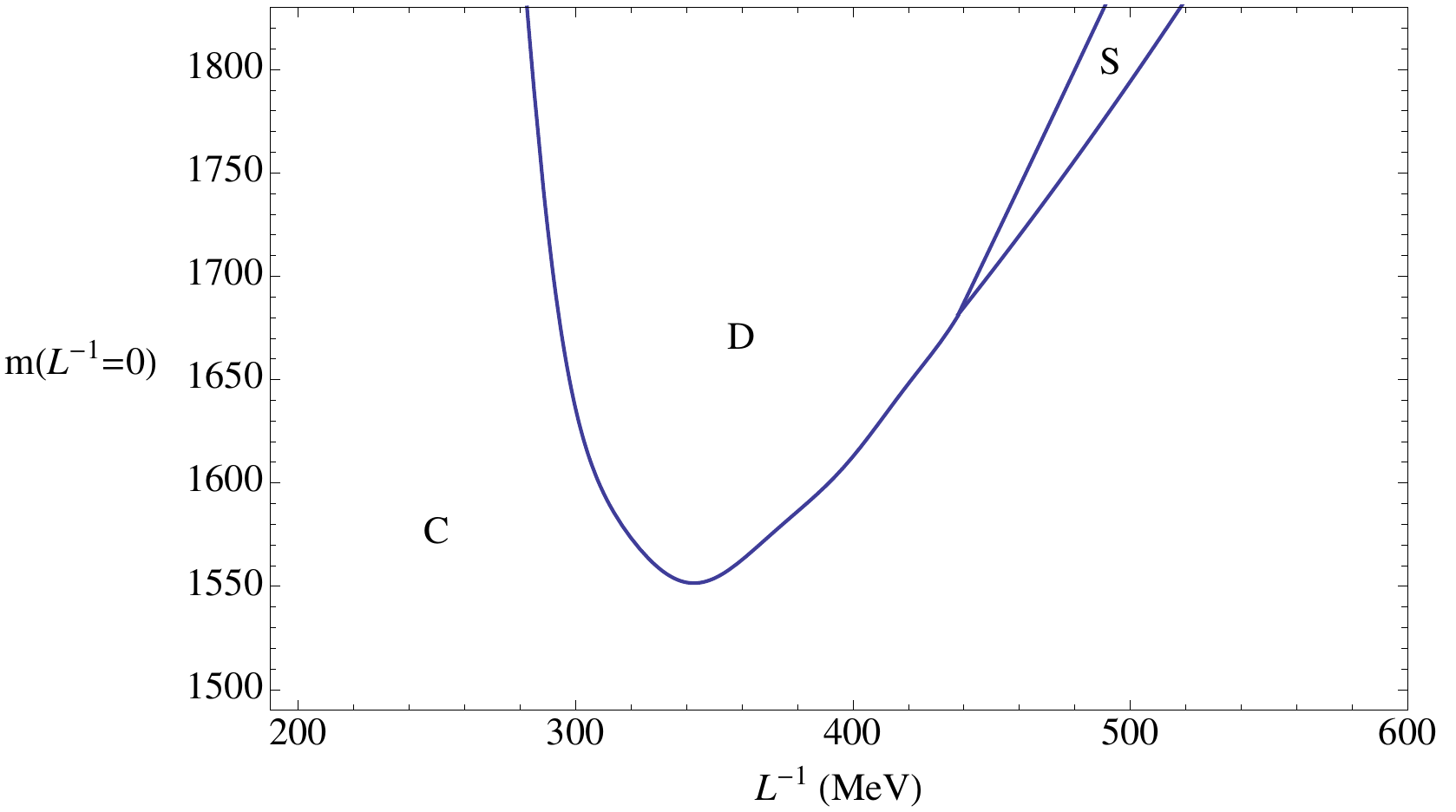}
\caption{The phase diagram for two-flavor
QCD with adjoint representation fermions with periodic boundary conditions
in terms of $L^{-1}$ and the constituent mass
at $L^{-1}=0$, $m(L^{-1}=0)$ for one choice of $\Lambda$. 
C, D and S refer to the confined phase, deconfined phase,
and skewed phase respectively. The constituent
mass is given in MeV.}
\label{fig:PhaseDiagram_Fit_TvsMass}
\end{figure}

PNJL models have a larger set
of parameters available, namely $\left\{ M,m_{0},\kappa,\Lambda \right\}$, than
in  adjoint $SU(N)$ gauge theories, 
which have only $m_{0}$ and $\Lambda_{adj}$. To
the extent that a PNJL model faithfully reproduces
the physics of adjoint $SU(N)$, we can think of adjoint $SU(N)$ as giving
a two-dimensional surface in the four-dimensional space of PNJL parameters.
For the case $m_{0}=0$, where the Lagrangians are chirally symmetric,
we obtain a line through a three-dimensional space. This motivates
us to consider the phase structure of our PNJL model in the 
$L^{-1}-\kappa$ plane with $\Lambda$ and $M$ fixed. 
Note however, that the PNJL
model will not break chiral symmetry at $L^{-1}=0$ unless 
$\kappa>\kappa_{c}=\pi^2/8\simeq 1.234$.
\begin{figure}
\includegraphics[width=5in]{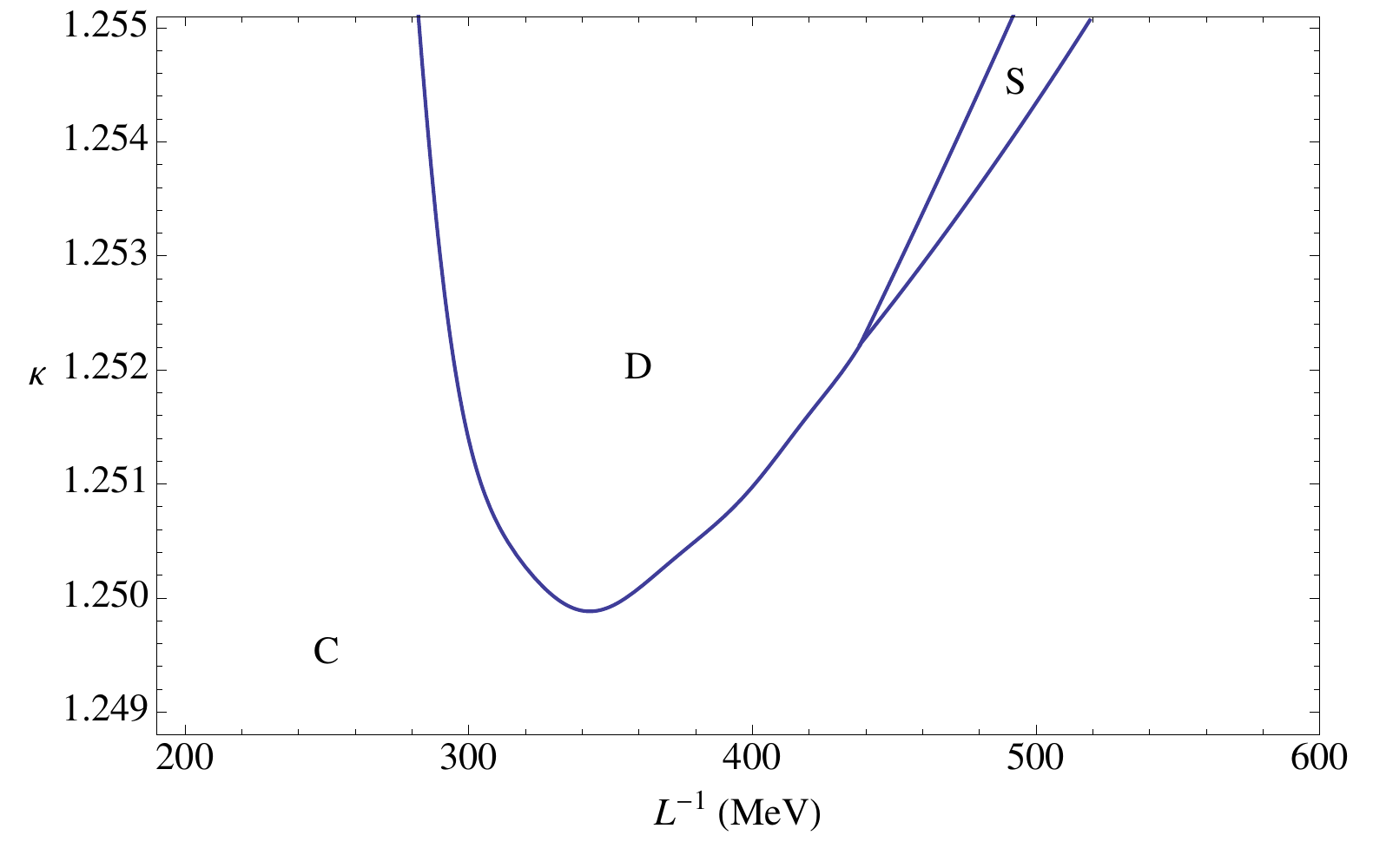}
\caption{The phase diagram for two-flavor
QCD with adjoint representation fermions with periodic boundary conditions
in the $L^{-1}$-$\kappa$ plane for one choice of $\Lambda$.
C, D and S refer to the confined phase, deconfined phase,
and skewed phase respectively.}
\label{fig:PhaseDiagram_Fit}
\end{figure}
In Figure \ref{fig:PhaseDiagram_Fit}, 
we show the phase diagram in the $L^{-1}-\kappa$ plane,
with $m_{0}=0$ and $M$ and $\Lambda$ as before, obtained by numerically
minimizing $V_{eff}$. For most values of $\kappa$ larger than $\kappa_{c}$,
the confined large-$L$ phase and the reconfined phase at 
small $L$ are separated by three phase transitions. There is a transition
from the confined phase to the deconfined phase, then another transition
from the deconfined phase to the skewed phase, followed by a third
transition from the skewed phase to the reconfined phase. All of these
transitions are characterized by abrupt changes in $Tr_{F}P$, while
the chiral order parameter shows only a slow decrease with increasing
temperature. However, there is a narrow range of $\kappa$ between
approximately $1.250$ and $\kappa_{c}\simeq 1.234$ where confinement holds at all
temperatures, and chiral symmetry remains broken. In this extended
phase diagram, the confined and reconfined regions are smoothly connected.
Although this connection appears only for small range of $\kappa$
values, the corresponding range of constituent mass values is not
necessarily small. In figure \ref{fig:PhaseDiagram_Fit_TvsMass}, we show the same phase diagram,
now plotted with $m(L^{-1}=0)$ replacing $\kappa$. 

There is a remaining puzzle associated with chiral symmetry restoration.
If $\kappa<\kappa_{c}$, there is no chiral symmetry breaking,
the adjoint fermions are always light, and $Tr_{F}P=0$ for all values
of $L^{-1}$. The region $\kappa<\kappa_{c}$, is separated from the
confined phase with $\kappa>\kappa_{c}$ by a chiral transition in
which $m>0$ for $\kappa>\kappa_{c}$. However, for $m_{0}$ light
but non-zero, there is no true chiral transition across $\kappa=\kappa_{c}$,
only a rapid crossover. The apparent existence within this model of
a confining phase with unbroken chiral symmetry must give us pause.
There is a long-standing argument due to Casher that suggests that
under very general circumstances
confining theories must also break chiral symmetry \cite{Casher:1979vw}.
Confinement without chiral symmetry breaking will also occur
in this model if $L$ is sufficiently small, even if $\kappa >\kappa_c$.
It is not clear if this is due to a defect in Casher's argument,
a failure of the PNJL model, or something more subtle.

Figures \ref{fig:PhaseDiagram_Fit}
and  \ref{fig:PhaseDiagram_Fit_TvsMass}
show that the small-$L$ and large-$L$ confining
phases are connected, at least within the PNJL model.
A similar smooth connection between the two regions has
been observed in lattice simulations using a deformation of the pure
gauge action in which a $Tr_{A}P$ term is added to the action \cite{Myers:2007vc}.
However, the connection between the
large-$L$ and small-$L$ regions
seen here in the PNJL model may not appear
in adjoint QCD without additional terms in the action.
Our results bear directly on the recent work by Cossu and D'Elia 
\cite{Cossu:2009sq}, in which they performed lattice simulations of two-flavor
$SU(3)$ gauge theory with periodic adjoint fermions. The simulations
were carried out on $16^{3}\times4$ lattices at various values of
the dimensionless lattice parameters, the gauge coupling $\beta$
and the dimensionless lattice fermion mass $m_{l}$. The size of the
compact dimension $L$ in physical units is given by $L=4a$, where
$a$ is determined from $\beta$ and $m_{l}$ via the renormalization
group. At fixed $m_{l}$, larger values of $\beta$ correspond to smaller
values of $L$. In these simulations, clear evidence was found for
confined, deconfined, skewed, and finally again confined phases as
$\beta$ was increased, corresponding to shrinking $L$. As $m_{l}$
was decreased, the critical values of $\beta$ at which the three
transitions occurred moved closer together. However, an extrapolation
to $m_{l}=0$ indicates that the two confined regions remain
separated as $m_{l}$ goes to zero. The deconfined phase clearly persists
in this limit, but the skewed phase  may or may not disappear when $m_{l}=0$.
All lattice simulations on finite lattices
have lower limits on the Lagrangian mass $m_{0}$, so it is \emph{a
priori} possible that the two confining regions seen
in the simulations are connected, but for a very
small range of $m_{0}$.
The phase structure seen in these lattice simulations
can also be obtained from our PNJL model if
the chiral limit, $m_0 = 0$, lies near
the triple point where the confined, deconfined,
and skewed phases meet.
For example, if we fix $m(L^{-1}=0)$ by adjusting $\kappa$,
and then decrease $L$,
we will obtain either one, three, or four phases,
as may be seen from Fig. \ref{fig:PhaseDiagram_Fit_TvsMass}.
The figure shows that the scale over which $m(L^{-1}=0)$
must vary in order to obtain the different phase structures
is quite small, on the order of $150\, MeV$, compared to
the scale of $m(L^{-1}=0)$ itself. 

\section{Conclusions}

We have extended the PNJL treatment of $SU(3)$ gauge theories to
the case of adjoint fermions with periodic boundary conditions on
$R^{3}\times S^{1}$. This class of models can have $Z(N)$ symmetry
for small $L$ provided the constituent mass of the fermions is sufficiently
light. The constituent mass rather than the Lagrangian mass is relevant
because the constituent mass determines the fermionic coupling to
the Polyakov loop. Previous analyses implicitly assumed that the constituent
mass could be set arbitrarily, ignoring the effects of chiral symmetry
breaking. 
Because PNJL models give a phenomenological description of 
both deconfinement and chiral symmetry restoration, 
they are natural tools for exploring
adjoint fermions with periodic boundary conditions.

We have shown
that a simple model reproduces the
known successes of PNJL models for fundamental fermions
while at the same time reproducing
the expected behavior at high temperatures
needed with adjoint fermions. 
The large separation between the deconfinement
transition and the chiral symmetry restoration transition for adjoint
fermion theories with antiperiodic boundary conditions requires a PNJL
model which reproduces the behavior of the pure gauge theory to much
smaller values of $L$ than have been considered before. 
We have extensively studied the case of the $SU(3)$ gauge
theory with two fermions degenerate in mass. This two-flavor case
is particularly simple to work with, because the effective
potential can be written as a function of a single coupling constant.

The results for our $SU(3)$ PNJL model with two flavors of periodic
adjoint Dirac fermions can be summarized in terms of two parameters: $\kappa$,
which represents the dimensionless strength of four-fermion interaction,
and $m_{0}$, the Lagrangian mass of the fermions. As with all NJL models,
there is a value $\kappa_{c}$ below which chiral symmetry breaking
does not take place, even as $L \rightarrow \infty$. 
Because chiral symmetry is broken
at $L^{-1}=0$ in the corresponding gauge theory, $\kappa$ must be
taken greater than $\kappa_{c}$ at large $L$ in order to reproduce
the behavior of the associated gauge theory. 
For any fixed values of $\kappa$
and $m_{0}$, there will be a value of $L$ below which the system
will be in the confined phase. In most of those cases, the system
will pass through the confined, deconfined, and skewed phases as $L$
is decreased, before the confined phase is regained.
There is a narrow range of parameters where only the deconfined
phase lies between the two confining regions.
If $m_{0}$ is
set to zero, there is a small region in the $L^{-1}-\kappa$ plane, lying
above $\kappa_{c}$, that connects the large-$L$ and small-$L$ confined
regions. Because the largest contribution to the constituent mass
$m$ is from chiral symmetry breaking, this behavior will persist
for some small range of non-zero $m_{0}$. Thus there is a single
confining region, accessible in principle in lattice simulations via
a gauge theory to which additional four-fermion terms have been added.
As we have seen, our results for the PNJL model
are completely compatible with the lattice
simulations of Cossu and D'Elia 
\cite{Cossu:2009sq}, in which they performed lattice simulations of two-flavor
$SU(3)$ gauge theory with periodic adjoint fermions.
The phase diagram obtained from
their simulations can easily be obtained
from the PNJL model with the appropriate
choice of parameters.
With more input from lattice simulations
of adjoint fermions, particularly
of basic hadronic properties for large $L$,
we could match the behavior of the PNJL
model more closely to the gauge theory.

Adjoint fermions with periodic boundary conditions provide a local
field theory that induces confinement in a small-$L$ region 
where semiclassical
arguments are valid. It appears
likely that the two confinement regions, small-$L$ and large-$L$,
can be connected if additional
terms, corresponding to irrelevant operators, are added to the action.
This may have implications for the use of periodic adjoint fermions
in achieving volume independence in the large-$N$ limit 
\cite{Unsal:2008ch,Bringoltz:2009mi,Bringoltz:2009kb,Poppitz:2009fm}, 
which requires unbroken $Z(N)$ symmetry.
Perhaps even more important is  our newly-gained ability
to understand confinement in a four-dimensional
gauge theory, albeit under somewhat exotic and unexpected
circumstances.

\begin{acknowledgments}
The authors thank the U.S. Department of Energy
for financial support.
\end{acknowledgments}


\bibliography{SU3_Adjoint_PBC_ms-Hiro}

\begin{thebibliography}{29}
\expandafter\ifx\csname natexlab\endcsname\relax\def\natexlab#1{#1}\fi
\expandafter\ifx\csname bibnamefont\endcsname\relax
  \def\bibnamefont#1{#1}\fi
\expandafter\ifx\csname bibfnamefont\endcsname\relax
  \def\bibfnamefont#1{#1}\fi
\expandafter\ifx\csname citenamefont\endcsname\relax
  \def\citenamefont#1{#1}\fi
\expandafter\ifx\csname url\endcsname\relax
  \def\url#1{\texttt{#1}}\fi
\expandafter\ifx\csname urlprefix\endcsname\relax\def\urlprefix{URL }\fi
\providecommand{\bibinfo}[2]{#2}
\providecommand{\eprint}[2][]{\url{#2}}

\bibitem[{\citenamefont{Unsal}(2008)}]{Unsal:2007vu}
\bibinfo{author}{\bibfnamefont{M.}~\bibnamefont{Unsal}},
  \bibinfo{journal}{Phys. Rev. Lett.} \textbf{\bibinfo{volume}{100}},
  \bibinfo{pages}{032005} (\bibinfo{year}{2008}), \eprint{0708.1772}.

\bibitem[{\citenamefont{Myers and Ogilvie}(2008)}]{Myers:2007vc}
\bibinfo{author}{\bibfnamefont{J.~C.} \bibnamefont{Myers}} \bibnamefont{and}
  \bibinfo{author}{\bibfnamefont{M.~C.} \bibnamefont{Ogilvie}},
  \bibinfo{journal}{Phys. Rev.} \textbf{\bibinfo{volume}{D77}},
  \bibinfo{pages}{125030} (\bibinfo{year}{2008}), \eprint{0707.1869}.

\bibitem[{\citenamefont{Eguchi and Kawai}(1982)}]{Eguchi:1982nm}
\bibinfo{author}{\bibfnamefont{T.}~\bibnamefont{Eguchi}} \bibnamefont{and}
  \bibinfo{author}{\bibfnamefont{H.}~\bibnamefont{Kawai}},
  \bibinfo{journal}{Phys. Rev. Lett.} \textbf{\bibinfo{volume}{48}},
  \bibinfo{pages}{1063} (\bibinfo{year}{1982}).

\bibitem[{\citenamefont{Kovtun et~al.}(2007)\citenamefont{Kovtun, Unsal, and
  Yaffe}}]{Kovtun:2007py}
\bibinfo{author}{\bibfnamefont{P.}~\bibnamefont{Kovtun}},
  \bibinfo{author}{\bibfnamefont{M.}~\bibnamefont{Unsal}}, \bibnamefont{and}
  \bibinfo{author}{\bibfnamefont{L.~G.} \bibnamefont{Yaffe}},
  \bibinfo{journal}{JHEP} \textbf{\bibinfo{volume}{06}}, \bibinfo{pages}{019}
  (\bibinfo{year}{2007}), \eprint{hep-th/0702021}.

\bibitem[{\citenamefont{Bringoltz and Sharpe}(2008)}]{Bringoltz:2008av}
\bibinfo{author}{\bibfnamefont{B.}~\bibnamefont{Bringoltz}} \bibnamefont{and}
  \bibinfo{author}{\bibfnamefont{S.~R.} \bibnamefont{Sharpe}},
  \bibinfo{journal}{Phys. Rev.} \textbf{\bibinfo{volume}{D78}},
  \bibinfo{pages}{034507} (\bibinfo{year}{2008}), \eprint{0805.2146}.

\bibitem[{\citenamefont{Gross et~al.}(1981)\citenamefont{Gross, Pisarski, and
  Yaffe}}]{Gross:1980br}
\bibinfo{author}{\bibfnamefont{D.~J.} \bibnamefont{Gross}},
  \bibinfo{author}{\bibfnamefont{R.~D.} \bibnamefont{Pisarski}},
  \bibnamefont{and} \bibinfo{author}{\bibfnamefont{L.~G.} \bibnamefont{Yaffe}},
  \bibinfo{journal}{Rev. Mod. Phys.} \textbf{\bibinfo{volume}{53}},
  \bibinfo{pages}{43} (\bibinfo{year}{1981}).

\bibitem[{\citenamefont{Weiss}(1981)}]{Weiss:1980rj}
\bibinfo{author}{\bibfnamefont{N.}~\bibnamefont{Weiss}},
  \bibinfo{journal}{Phys. Rev.} \textbf{\bibinfo{volume}{D24}},
  \bibinfo{pages}{475} (\bibinfo{year}{1981}).

\bibitem[{\citenamefont{Meisinger and Ogilvie}(2002)}]{Meisinger:2001fi}
\bibinfo{author}{\bibfnamefont{P.~N.} \bibnamefont{Meisinger}}
  \bibnamefont{and} \bibinfo{author}{\bibfnamefont{M.~C.}
  \bibnamefont{Ogilvie}}, \bibinfo{journal}{Phys. Rev.}
  \textbf{\bibinfo{volume}{D65}}, \bibinfo{pages}{056013}
  (\bibinfo{year}{2002}), \eprint{hep-ph/0108026}.

\bibitem[{\citenamefont{Myers and Ogilvie}(2009)}]{Myers:2009df}
\bibinfo{author}{\bibfnamefont{J.~C.} \bibnamefont{Myers}} \bibnamefont{and}
  \bibinfo{author}{\bibfnamefont{M.~C.} \bibnamefont{Ogilvie}},
  \bibinfo{journal}{JHEP} \textbf{\bibinfo{volume}{07}}, \bibinfo{pages}{095}
  (\bibinfo{year}{2009}), \eprint{0903.4638}.

\bibitem[{\citenamefont{Davies et~al.}(1999)\citenamefont{Davies, Hollowood,
  Khoze, and Mattis}}]{Davies:1999uw}
\bibinfo{author}{\bibfnamefont{N.~M.} \bibnamefont{Davies}},
  \bibinfo{author}{\bibfnamefont{T.~J.} \bibnamefont{Hollowood}},
  \bibinfo{author}{\bibfnamefont{V.~V.} \bibnamefont{Khoze}}, \bibnamefont{and}
  \bibinfo{author}{\bibfnamefont{M.~P.} \bibnamefont{Mattis}},
  \bibinfo{journal}{Nucl. Phys.} \textbf{\bibinfo{volume}{B559}},
  \bibinfo{pages}{123} (\bibinfo{year}{1999}), \eprint{hep-th/9905015}.

\bibitem[{\citenamefont{Davies et~al.}(2003)\citenamefont{Davies, Hollowood,
  and Khoze}}]{Davies:2000nw}
\bibinfo{author}{\bibfnamefont{N.~M.} \bibnamefont{Davies}},
  \bibinfo{author}{\bibfnamefont{T.~J.} \bibnamefont{Hollowood}},
  \bibnamefont{and} \bibinfo{author}{\bibfnamefont{V.~V.} \bibnamefont{Khoze}},
  \bibinfo{journal}{J. Math. Phys.} \textbf{\bibinfo{volume}{44}},
  \bibinfo{pages}{3640} (\bibinfo{year}{2003}), \eprint{hep-th/0006011}.

\bibitem[{\citenamefont{Gocksch and Ogilvie}(1985)}]{Gocksch:1984yk}
\bibinfo{author}{\bibfnamefont{A.}~\bibnamefont{Gocksch}} \bibnamefont{and}
  \bibinfo{author}{\bibfnamefont{M.}~\bibnamefont{Ogilvie}},
  \bibinfo{journal}{Phys. Rev.} \textbf{\bibinfo{volume}{D31}},
  \bibinfo{pages}{877} (\bibinfo{year}{1985}).

\bibitem[{\citenamefont{Fukushima}(2004)}]{Fukushima:2003fw}
\bibinfo{author}{\bibfnamefont{K.}~\bibnamefont{Fukushima}},
  \bibinfo{journal}{Phys. Lett.} \textbf{\bibinfo{volume}{B591}},
  \bibinfo{pages}{277} (\bibinfo{year}{2004}), \eprint{hep-ph/0310121}.

\bibitem[{\citenamefont{Karsch and Lutgemeier}(1999)}]{Karsch:1998qj}
\bibinfo{author}{\bibfnamefont{F.}~\bibnamefont{Karsch}} \bibnamefont{and}
  \bibinfo{author}{\bibfnamefont{M.}~\bibnamefont{Lutgemeier}},
  \bibinfo{journal}{Nucl. Phys.} \textbf{\bibinfo{volume}{B550}},
  \bibinfo{pages}{449} (\bibinfo{year}{1999}), \eprint{hep-lat/9812023}.

\bibitem[{\citenamefont{Engels et~al.}(2005)\citenamefont{Engels, Holtmann, and
  Schulze}}]{Engels:2005te}
\bibinfo{author}{\bibfnamefont{J.}~\bibnamefont{Engels}},
  \bibinfo{author}{\bibfnamefont{S.}~\bibnamefont{Holtmann}}, \bibnamefont{and}
  \bibinfo{author}{\bibfnamefont{T.}~\bibnamefont{Schulze}},
  \bibinfo{journal}{Nucl. Phys.} \textbf{\bibinfo{volume}{B724}},
  \bibinfo{pages}{357} (\bibinfo{year}{2005}), \eprint{hep-lat/0505008}.

\bibitem[{\citenamefont{Klevansky}(1992)}]{Klevansky:1992qe}
\bibinfo{author}{\bibfnamefont{S.~P.} \bibnamefont{Klevansky}},
  \bibinfo{journal}{Rev. Mod. Phys.} \textbf{\bibinfo{volume}{64}},
  \bibinfo{pages}{649} (\bibinfo{year}{1992}).

\bibitem[{\citenamefont{Hatsuda and Kunihiro}(1994)}]{Hatsuda:1994pi}
\bibinfo{author}{\bibfnamefont{T.}~\bibnamefont{Hatsuda}} \bibnamefont{and}
  \bibinfo{author}{\bibfnamefont{T.}~\bibnamefont{Kunihiro}},
  \bibinfo{journal}{Phys. Rept.} \textbf{\bibinfo{volume}{247}},
  \bibinfo{pages}{221} (\bibinfo{year}{1994}), \eprint{hep-ph/9401310}.

\bibitem[{\citenamefont{Cossu and D'Elia}(2009)}]{Cossu:2009sq}
\bibinfo{author}{\bibfnamefont{G.}~\bibnamefont{Cossu}} \bibnamefont{and}
  \bibinfo{author}{\bibfnamefont{M.}~\bibnamefont{D'Elia}},
  \bibinfo{journal}{JHEP} \textbf{\bibinfo{volume}{07}}, \bibinfo{pages}{048}
  (\bibinfo{year}{2009}), \eprint{0904.1353}.

\bibitem[{\citenamefont{Unsal}(2009)}]{Unsal:2007jx}
\bibinfo{author}{\bibfnamefont{M.}~\bibnamefont{Unsal}},
  \bibinfo{journal}{Phys. Rev.} \textbf{\bibinfo{volume}{D80}},
  \bibinfo{pages}{065001} (\bibinfo{year}{2009}), \eprint{0709.3269}.

\bibitem[{\citenamefont{Brower et~al.}(1994)\citenamefont{Brower, Shen, and
  Tan}}]{Brower:1994sw}
\bibinfo{author}{\bibfnamefont{R.~C.} \bibnamefont{Brower}},
  \bibinfo{author}{\bibfnamefont{Y.}~\bibnamefont{Shen}}, \bibnamefont{and}
  \bibinfo{author}{\bibfnamefont{C.-I.} \bibnamefont{Tan}},
  \bibinfo{journal}{Nucl. Phys. Proc. Suppl.} \textbf{\bibinfo{volume}{34}},
  \bibinfo{pages}{210} (\bibinfo{year}{1994}), \eprint{hep-lat/9403011}.

\bibitem[{\citenamefont{Brower et~al.}(1995)\citenamefont{Brower, Orginos, and
  Tan}}]{Brower:1995vf}
\bibinfo{author}{\bibfnamefont{R.~C.} \bibnamefont{Brower}},
  \bibinfo{author}{\bibfnamefont{K.}~\bibnamefont{Orginos}}, \bibnamefont{and}
  \bibinfo{author}{\bibfnamefont{C.~I.} \bibnamefont{Tan}},
  \bibinfo{journal}{Nucl. Phys. Proc. Suppl.} \textbf{\bibinfo{volume}{42}},
  \bibinfo{pages}{42} (\bibinfo{year}{1995}), \eprint{hep-lat/9501026}.

\bibitem[{\citenamefont{Kogut et~al.}(1998)\citenamefont{Kogut, Lagae, and
  Sinclair}}]{Kogut:1998rg}
\bibinfo{author}{\bibfnamefont{J.~B.} \bibnamefont{Kogut}},
  \bibinfo{author}{\bibfnamefont{J.~F.} \bibnamefont{Lagae}}, \bibnamefont{and}
  \bibinfo{author}{\bibfnamefont{D.~K.} \bibnamefont{Sinclair}},
  \bibinfo{journal}{Phys. Rev.} \textbf{\bibinfo{volume}{D58}},
  \bibinfo{pages}{034504} (\bibinfo{year}{1998}), \eprint{hep-lat/9801019}.

\bibitem[{\citenamefont{Meisinger et~al.}(2002)\citenamefont{Meisinger, Miller,
  and Ogilvie}}]{Meisinger:2001cq}
\bibinfo{author}{\bibfnamefont{P.~N.} \bibnamefont{Meisinger}},
  \bibinfo{author}{\bibfnamefont{T.~R.} \bibnamefont{Miller}},
  \bibnamefont{and} \bibinfo{author}{\bibfnamefont{M.~C.}
  \bibnamefont{Ogilvie}}, \bibinfo{journal}{Phys. Rev.}
  \textbf{\bibinfo{volume}{D65}}, \bibinfo{pages}{034009}
  (\bibinfo{year}{2002}), \eprint{hep-ph/0108009}.

\bibitem[{\citenamefont{Meisinger and Ogilvie}(2009)}]{Meisinger:2009ne}
\bibinfo{author}{\bibfnamefont{P.~N.} \bibnamefont{Meisinger}}
  \bibnamefont{and} \bibinfo{author}{\bibfnamefont{M.~C.}
  \bibnamefont{Ogilvie}} (\bibinfo{year}{2009}), \eprint{0905.3577}.

\bibitem[{\citenamefont{Casher}(1979)}]{Casher:1979vw}
\bibinfo{author}{\bibfnamefont{A.}~\bibnamefont{Casher}},
  \bibinfo{journal}{Phys. Lett.} \textbf{\bibinfo{volume}{B83}},
  \bibinfo{pages}{395} (\bibinfo{year}{1979}).

\bibitem[{\citenamefont{Unsal and Yaffe}(2008)}]{Unsal:2008ch}
\bibinfo{author}{\bibfnamefont{M.}~\bibnamefont{Unsal}} \bibnamefont{and}
  \bibinfo{author}{\bibfnamefont{L.~G.} \bibnamefont{Yaffe}},
  \bibinfo{journal}{Phys. Rev.} \textbf{\bibinfo{volume}{D78}},
  \bibinfo{pages}{065035} (\bibinfo{year}{2008}), \eprint{0803.0344}.

\bibitem[{\citenamefont{Bringoltz}(2009)}]{Bringoltz:2009mi}
\bibinfo{author}{\bibfnamefont{B.}~\bibnamefont{Bringoltz}},
  \bibinfo{journal}{JHEP} \textbf{\bibinfo{volume}{06}}, \bibinfo{pages}{091}
  (\bibinfo{year}{2009}), \eprint{0905.2406}.

\bibitem[{\citenamefont{Bringoltz and Sharpe}(2009)}]{Bringoltz:2009kb}
\bibinfo{author}{\bibfnamefont{B.}~\bibnamefont{Bringoltz}} \bibnamefont{and}
  \bibinfo{author}{\bibfnamefont{S.~R.} \bibnamefont{Sharpe}}
  (\bibinfo{year}{2009}), \eprint{0906.3538}.

\bibitem[{\citenamefont{Poppitz and Unsal}(2009)}]{Poppitz:2009fm}
\bibinfo{author}{\bibfnamefont{E.}~\bibnamefont{Poppitz}} \bibnamefont{and}
  \bibinfo{author}{\bibfnamefont{M.}~\bibnamefont{Unsal}}
  (\bibinfo{year}{2009}), \eprint{0911.0358}.

\end{thebibliography}

\end{document}